\documentclass[twocolumn,showpacs,prd,amsmath,amssymb,superscriptaddress]{revtex4}

\usepackage{graphicx}
\usepackage{dcolumn}
\usepackage{bm}

\begin{document}

\title{Quantum amplification effect in a horizon fluctuations}

\author{Mohammad H. Ansari}
\email{mhansari@math.uwaterloo.ca}
\affiliation{Department of Combinatorics and Optimization, University of Waterloo, Waterloo, Ontario, Canada N2L 3G1 }%
\date{\today}
\newcommand{\beq}{\begin{equation}}
\newcommand{\eeq}{\end{equation}}
\newcommand{\barr}{\begin{array}}
\newcommand{\earr}{\end{array}}
\newcommand{\ssz}{\scriptsize}
\newcommand{\amin}{a_{\mathrm{min}}}
\newcommand{\zmin}{\zeta_{\mathrm{min}}}

\begin{abstract}
The appearance of a few unevenly- spaced bright flashes of light on top of Hawking radiation is the sign of the amplification effect in black hole horizon fluctuations. Previous studies on this problem suffer from the lack of considering all emitted photons in the theoretical spectroscopy of these fluctuations.   In this paper, we include all of the physical transition weights and present a consistent intensity formula. This modifies a black hole radiation pattern.
\end{abstract}

\maketitle



\section{Introduction}

Since the time of predicting black hole radiation by Hawking \cite{Hawking:1974sw}, it has been argued this radiation may not become observable due to its low intensity in stable holes \cite{Frampton:2005fk}. Nevetheless, the derivation of this radiation is incomplete and relies on a fixed nonperturbing background whose quantum properties of gravity are not present.

Semiclassical quantization analysis predicts a discrete area for rotating uncharged holes \cite{Gour}, charged holes \cite{conti}, Kerr and extremal Kerr holes \cite{das}, and some others  \cite{{others},{oth2},{oth3},{oth4}}. A large class of these studies predict a discrete spectrum of area of the form $A \sim \sqrt{n}$, which is obviously unevenly- spaced.  Loop quantum gravity also supports this discreteness on any surface \cite{Ansari:2006vg}.  Therefore one can expect this area quantization to change the Hawking radiation picture due to the fluctuations of the horizon area.

Based on the heuristic proposal of an evenly- spaced quantization of area by Bekenstein in \cite{Bekensteinoriginal}, Bekenstein and Mukhanov worked out the effect of fluctuations of horizon area in \cite{{Bekenstein:1995ju},{bek1},{bek2}}. Because the area of a black hole surface is connected to the black hole mass, the black hole mass is likely to be quantized as well. The black hole mass decreases when it radiates; therefore its quantum of mass decreases by a finite value after one emission, similar to the way atoms decay. A consequence of this picture is that radiation is emitted at quantized frequencies, corresponding to the differences between energy levels. Therefore the version of quantum gravity they applied implies an evenly-spaced discrete emission spectrum for the black hole radiation on top of Hawking radiation, \cite{{bek1},{bek2}}.

Ansari in \cite{Ansari:2006vg} studied theories with unevenly-spaced quantum of area, such as those with area spectrum $A \sim \sqrt{n}$. The fluctuations of such a horizon results in a continuous spectrum; however at some discrete frequencies  an avalanche of many copies of a photon is generated by the hole while other frequencies remain single-copied. Therefore, the intensity is inhomogeneous such that the amplified modes are radiates in intense lines, similar to a laser system.  As a result the transitions between equidistant subsets amplify by the hole  and prevail in the radiated spectrum. The overlapping of all transitions produces an `unevenly-spaced' spectrum of amplified lines and only a few of these lines are the brightest ones.

Recently some attempts have been made to search for these lines in high energy observations \cite{khriplovich}. Therefore, it is crucially important to extract the spectroscopy of the highly amplified lines. We noticed in the previous studies on determining this spectroscopy \cite{{Ansari:2006vg},{Ansari:2006cx}} only a limited number of photons have been counted due to ignoring the transition weight. Here, we introduce this factor and derive the brightness intensity. Unlike the results presented in \cite{Ansari:2006vg} the intensity is completely independent of the Barbero-Immirzi parameter and scales with the frequency harmonic number in a power law. As a result, the spectroscopy profile of a black hole radiation is changed such that in the interval of $\omega/\omega_o <3 $ one can expect to observe a few unevenly-spaced lines with maximum intensity. The frequency scale $\omega_o$ is shown to depend on the black hole mass and to be independent of the Barbero-Immirzi parameter.

\section{Near horizon geometry}

A classical event horizon is a null boundary between two partitions of space-time by definition. This boundary is not locally defined, not even in time.  To define this boundary, one needs the information of the entire manifold.  Therefore other than expansions of geodesic congruences used in general relativity, a suitable local information flow definition is needed to define a black hole horizon locally. In canonical quantum gravity,
a definition is needed by which one looks at a place in space and says those photons that are reaching to us must come from a spatial
slice that intersects a space-time horizon. Such local definitions
are those of apparent, trapping, and dynamical horizons \cite{Ashtekar:1997yu}.

These space-time horizons
are \emph{not} necessarily null. They would be so if we have 1) vacuum
and 2) the absence of gravitational radiation. Vacuum can easily be
achieved for  spin networks, but we cannot prevent the local
gravitational degrees of freedom to be excited in the neighborhood
of a space-time horizon.  With this gravitational radiation across
the horizon and with positive energy conditions (or vacuum) the
horizon will be \emph{spacelike} rather than null.

Moreover, the
energy conditions in quantum gravity could not be taken for granted,
even for semiclassical states, as long as violations occur on small
length scales. Thus, quantum space-time horizons can become
even \emph{timelike} with a two-way information transfer.

In the lack of energy  positivity and the presence of gravitational radiations, the extension of an event horizon in quantum scales can no longer be considered as being null. In other words, in the vicinity of a black hole one \emph{cannot} restrict the quantum fluctuations of the horizon area into the null boundary subset of area $A \sim \sqrt{j(j+1)}$ (for positive half-integer $j$), which scales approximately as an integer number, \cite{{Ashtekar:1997yu},{Rovelli}}.

\begin{figure}
  \includegraphics[width=4cm]{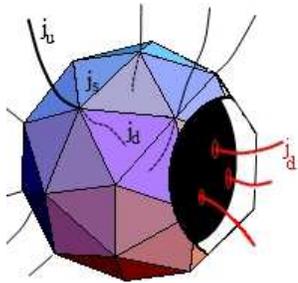}\\
  \caption{A quantum black hole horizon in the vicinity of a null boundary (the black sphere).}\label{Fig. blackhole}
\end{figure}

Instead, one may consider the fluctuations of black hole horizon into non-null surfaces in the vicinity of a classical null boundary. Figure (\ref{Fig. blackhole}) represents the null boundary in black sphere and non-null surfaces as a cover shell in its vicinity. The geometry of this shell can be connected into exterior and interior space-time sectors via  the out-coming and ingoing edges, respectively. There are also edges residing on the surface as well. The interior edges intersect the classical null boundary at the Chern-Simons punctures, which are degenerate wave functions \cite{Ashtekar:1997yu}.

 The quantum area at this surface is determined by the complete spectrum, obtained in loop quantum gravity as $A \sim \frac{1}{2}\sqrt{2 j_u (j_u+1) + 2j_d (j_d+1) - j_s(j_s+1)}$ conditioned to $|j_u-j_d| \leq j_s \leq j_u+j_d$, where $j_u$, $j_d$ and $j_s$ are half-integers and denote  local holonomy states in the upper, lower and surface sectors.  Using the number theory one can reformulate this complete spectrum of area in the form of  $A \sim \sqrt{n}$ for positive integer $n$.

In this paper our focus is on the class of theories that predicts area in the following form in the region infinitesimally near a black hole horizon:
\beq \label{eq. 1 A} a_n(\zeta) = a_0 \sqrt{n}\eeq
where $n$ is any natural number and we consider $a_o$ of the order of Planck's area. In Appendix A of \cite{Ansari:2006vg}, it is proven that a square root can be reformulated by integers as multipliers of square-free numbers $\zeta=\{1, 2, 3, 5, \cdots\}$, \footnote{In SU(2) group representation $\zeta$ turns out to be the Discriminant of Positive Definite Form numbers: $\{3,4,7,8,11,\cdots\}$.}. Therefore, we can reformulate (\ref{eq. 1 A}) into:
\beq \label{eq. universal A} a_n(\zeta) = a_0 \sqrt{\zeta}\ n\eeq

By fixing $\zeta$ in Eq. (\ref{eq. universal A}) an equidistant subset appears that is called a `generation'. The parameter $\zeta$ is called the
`generational number'. Let us also name the minimum generational number $\zmin$ and the corresponding minimal area $\amin$.  Note that the  term $\sqrt{\zeta}$ is an irrational number and is unique in a generation. Therefore the summation of any two quanta $a_{n_1}(\zeta_1)$ and $a_{n_2}(\zeta_2)$ of different generations ($\zeta_1 \neq \zeta_2$) belongs to none of the generations.

\section{Degeneracy}


The quantum states of the surfaces in the vicinity of a black hole horizon are degenerate states, \cite{Ansari:2006cx}. Large area patches in a generation can be exactly decomposed into smaller patches of the same generation.  For example, $a_n = n a_1 = (n-2) a_1 + a_2 =
\cdots$.  These cells are all completely distinguishable due to the presence of surface gravity excitations on a surface as discussed in \cite{Ansari:2006cx}. The multiplicity of area eigenvalue $a_n$ is therefore: $\Omega_n = g_n + g_{n-1}g_1 + \cdots + (g_1)^n$.
The dominant term in the sum belongs to the configuration
with the maximum number of the area cell $a_1$. Therefore the total
degeneracy of $a_n(\zeta)$ for $n \gg 1$ is: \beq \label{eq deg}
\Omega_n(\zeta) = g_1(\zeta)^n. \eeq

In the classical limits, the dominant configuration of a large
surface with respect to the Planck scale, is the nearest area level into the horizon area from
the generation of the minimal gap between levels; i. e. $A \approx n \amin$.  This
dominant degeneracy is $g_1(\zmin)^n$ and a kinematic entropy can be
associated with it proportional to the area:

\beq
\label{eq. Sbh}
S = A (\ln g_1(\zmin) / \amin).
\eeq

 Depending on the type of time evolution of the surface
this entropy may vanish, decrease, increase or remain unchanged in
the course of time. In other words, a classical surface
characterized by its area at each time slice possesses a finite
entropylike parameter. Space-time horizons as a class of physical
surfaces possess a nondecreasing entropy. In other words their
kinematical entropy in the course of time, due to the second black
hole thermodynamics law, is \emph{physical} entropy. We will show
in the next section such a horizon carries an entropy whose nature
is the total degeneracy of vacuum fluctuation modes responsible for
the thermal radiation of black hole. However, for the aim of this
note on the study of kinematics of fluctuations we disregard here
the issues of defining the Hamiltonian of a quantum horizon based on
spin foam, which is still an open problem.

\section{Fluctuations of a horizon}

Black holes are physical systems with nondecreasing entropies. Jacobson et.al. in \cite{Jacobson:2005kr} associates this entropy of a black hole with the region where the interior and exterior space-time sectors meet on the horizon, therefore we assume this entropy changes only minimally in the region infinitesimally near the hole. We assume during the latest steps of a surface falling into a horizon, after all gravitational perturbations are radiated away, the surface radiates energies that only depend on the event horizon area $ A = \frac{16 \pi G^2}{c^4} M^2$. Using (\ref{eq. universal A}) the energy fluctuations become $\delta M =  \frac{M_{Pl}}{ 8 M} \sqrt{\zeta} \delta n$.

This energy in the quantum picture is generated from the change of surface area and passing this geometrical energy into a correlated photon with this geometry. The transitions between area levels are classified into two classes: either 1) the generational transitions: those with initial and final levels at the same generation, or 2) the intergenerational transitions: those with initial and
final levels belonging in different generations.   The former case generates harmonic modes whose fundamental frequency is the minimal frequency of that generation. The intergenerational transitions produce inharmonic modes. The `fundamental frequency' of the harmonic frequencies is the jump
between two consecutive levels with frequency $\varpi (\zeta) =
( \sqrt{\zeta})\ \omega_o $, where $\omega_o:=\frac{c^3}{8GM}$ is the so-called `frequency scale'.

Let us estimate the range of these frequencies.   In a micro black hole of mass $10^{-18} M_{\odot}$ the event horizon area is about $10^{-29}\ (\mathrm{m}^2)$ and the temperature is of order $10^{11}$ K. The frequency scale is thus of the order of $\sim 10$ keV. Such a
typical hole has a horizon 40 order of magnitude larger than the
Planck length area. Therefore from each harmonic mode there are numerous
\emph{copies} emitted from different levels. Such modes are extremely amplified by the hole. On the other hand, since two levels of different generations have a unique difference, there
exists only \emph{one} copy from each inharmonic mode in all
possible transitions.

The difference between harmonic and inharmonic modes in their particle production populations make the harmonic modes to become heavily intensified by the hole, the so-called \emph{quantum amplification effect} (QAE). In other words, a black hole condensates its particle production mostly on harmonic modes. One important consequence of QAE is that the density matrix elements of
inharmonic modes can be regarded negligible at population. Therefore we can propose the
generational transitions matrix elements to be uniform.

\subsection{Transition weight}

In a transition down the level of a generation, there are two weight
factors: 1) the transition and 2) the population weights. Let us introduce them. Assume a hole of
large area $A$. When the hole jumps $f$ steps down the ladder of
levels in the generation $\zeta$, it emits a quanta of the frequency
$f\varpi(\zeta)$. This radiance energy could also be emitted
in the dominant configuration by radiating
$f\frac{a_1(\zeta)}{\amin}$ quanta of the fundamental frequency
$\varpi(\zmin)$. These two transitions, although they are of the same
frequency, appear with different possibilities. The
degeneracy ratio of these two is $\Omega(f
\varpi(\zeta))/\Omega(f\frac{a_1(\zeta)}{\amin} \varpi(\zmin))$ that
gives rise to the definition of \emph{transition weight}:
\beq \label{eq. theta}
\theta (\zeta,
f) = g_1(\zeta)^{f} g_1(\zmin)^{-f a_1(\zeta) / \amin}.\eeq

\subsection{Population weight}

The second
weight is the population one that comes from a different root. Because of QAE, from each harmonic frequency there
produced many copies in different levels on the generation. This
weight is in fact the number of possible quanta emitting from
different levels with the same frequency. It is easy to verify this
number is $N_{\varpi(\zeta)}-f+1 $ where $N_{\varpi(\zeta)}$ is the
number of copies from the fundamental frequency, and for near level
modes ($ f \ll N_{\varpi(\zeta)}$) it is $\frac{A}{a_1(\zeta)}$. We
absorb constants in normalization factors and the \emph{population weight}
near levels becomes:

\beq \label{eq. rho}
\rho(\zeta):= 1/\sqrt{\zeta}.
\eeq

Finally notice that within one generation when a space-time hole
jumps $f$ steps down the ladder of levels, the degeneracy decreases
by a factor of $g_1(\zeta)^{f}$.

\subsection{Fluctuations}

Having defined the relevant transitions for the case of an nonequidistant area spectrum, we use the same strategy used originally in \cite{Bekenstein:1995ju} to determine the relative intensities. The probability of an
$\omega_f(\zeta)$ emission from the area spectrum (\ref{eq. universal A})
is
\[C^{-1} \rho(\zeta) g_1(\zmin)^{-f
\sqrt{\zeta / \zmin}},\]
where $C$ is the normalization factor,
\footnote{From normalization $C = \sum_{\zeta} \rho(\zeta)/|1-
g_1(\zmin)^{-\sqrt{\zeta/\zmin}}|$.}. Let us denote this probability by $P(\omega_f(\zeta) | 1 )$ as a conditional probability. This keeps in mind the necessary information that should be taken care of when later on a sequence of emissions are considered.

At the moment when the black hole mass is $M_i$ it may proceed to decay to a state of lower mass $M_{i+1}$ and emits photons of energy $M_{i}-M_{i+1}$. This process that is stimulated by zero-point fluctuations of the vacuum energy, can be repeated in timely sequential order. Gerlach proposed this order originally for an incipient black hole in \cite{gerlach}.  Consider  successive emissions. The probability
of a string of sequential emissions is provided by multiplying the probability of each emission in the time-ordered manner.

First, note the
conditional probability of  $j$ emissions is $ \prod_{i=1}^j P(\omega_{f_i}(\zeta_i) | 1 )$.
The probability of the sequences to include $k$ emissions out of $j$
to be of the frequency $\omega_{f^*}(\zeta^*)$ (in no matter what
order) while the rest of accompanying emissions are of any value
except this frequency, is

\begin{widetext}
\[P \left( k,\ \omega_{f^*}\left(\zeta^*\right);\ \left\{ \omega_{f_1}\left(\zeta_1\right), \cdots \right\} | j \right) = \left(^j_k\right) \left[ P(\omega_{f^*}(\zeta^*) | 1) \right]^k \times \prod_{i=1; \ \zeta \neq \zeta^*}^{j-k} P(\omega_{f_i}(\zeta_i) | 1 ).\]
\end{widetext}

 The accompanying
 modes are allowed to accept any frequency except
 $\omega_{f^*}(\zeta^*)$ and therefore the probabilities of any accompanying
frequency should sum. From the definition of $C$, it is easy to find
out in each sum over accompanying modes instead of $\sum_{\omega
\neq \omega^*} P(\omega_{f_i}(\zeta_i) | 1 )$ we can replace $C -
P(\omega_{f^*}(\zeta^*) | 1 )$ that simplifies the probability to become

\begin{widetext}
\beq \label{eq. pkwstarbh}
P(k, \omega_{f^*}(\zeta^*)|j) =
 \left(_{j}^{k}\right) \
 \left[ P(\omega_{f^*}(\zeta^*) | 1 )\right]^k \ \
\left[C-P(\omega_{f^*}(\zeta^*) | 1 )\right]^{j-k}.
\eeq
\end{widetext}

Second, we need to determine what is the probability of a time-ordered sequence. The probabilities of zero and one jump (of no matter
what frequency) in the time interval $\Delta t$ are $P_{\Delta
t}(0)$ and $P_{\Delta t}(1)$, respectively. In the time interval $2
\Delta t$, the probabilities of zero, one, and two jumps are
$P_{\Delta t}(0)^2$, $2 P_{\Delta t}(0) P_{\Delta t}(1)$, and $2
P_{\Delta t}(0) P_{\Delta t} (2) + P_{\Delta t}(1)^2$, respectively.
By induction this is found for higher number of jumps in an interval
and for a longer time. A general solution for the equations of the
probability of $j$ time-ordered decays in an interval of time
$\Delta t$ is $P_{\Delta t}(j) = \frac{1}{j!} (\frac{\Delta
t}{\tau})^j \exp(\frac{\Delta t}{\tau})$. Multiplying this
probability with $P(k, \omega_{f^*}(\zeta^*)|j)$ and then summing
over all sequence dimensions $j \geq k$, it is easy to manipulate
the total probability of $k$
 emissions with frequency $\omega_{f^*}(\zeta^*)$ to be

 \[P_{\Delta t}(k, \omega_{f^*}(\zeta^*)) = \frac{1}{k!} (x^*_f)^k
 \exp(-x^*_f),\]
 where $x^*_f = \frac{\Delta t}{C \tau} \rho(\zeta^*) \
g_1(\zmin)^{-f \sqrt{\zeta / \zmin}}$. This indicates the
distribution of the number of quanta emitted in harmonic modes is
Poisson-like. By definition, the intensity of a mode is the total energy
 emitted in that frequency per unit time and area. The average number of
emissive quanta at a typical harmonic frequency is $\overline{k} =
\sum_{k=1}^{\infty} k P_{\Delta t}(k | \omega_{f}(\zeta)) $.
Calculating this summation gives rise to the intensity
\begin{equation}\label{eq. I}
  \frac{I(\omega_{f}(\zeta))}{I_o} =  fg_1(\zmin)^{-f \sqrt{\zeta / \zmin}}.
\end{equation}

\begin{figure}
\includegraphics[width=9cm]{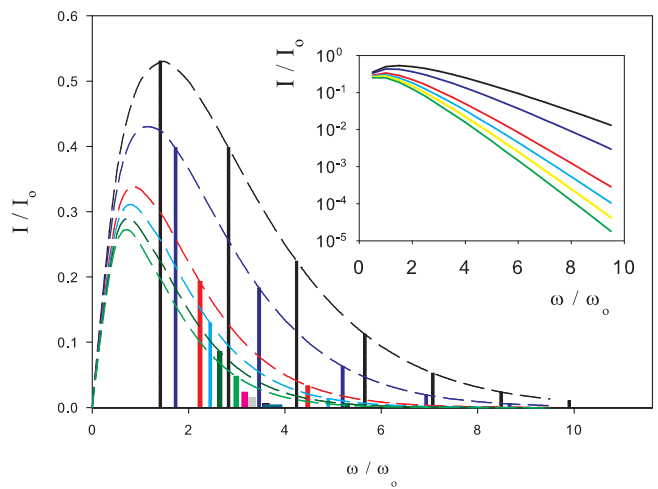}
\caption{The spectrum of highly amplified modes in the range of frequencies up to 10$\omega_o$. The intensity envelopes are in dashed lines. The inset indicates the intensity in log scales. The generations with smaller gaps appear with higher amplified intensities.}\label{fig2}\end{figure}

Comparing Eq. (\ref{eq. I}) with the intensity result in Ref. \cite{Ansari:2006cx} one can see some major differences. First, the intensity is independent of the Barbero-Immirzi parameter, while the intensity in Ref. \cite{Ansari:2006cx} due to the lack of the transition weight analysis gave rise to a formula that strictly depends on this parameter. Second,  the intensity scales as a  power law with the harmonic number.   Third, the intensity suppresses as a power law with $\sqrt{\zeta}$, which provides a very high suppression of high generations and harmonics.

These lines will not blend because they are exactly similar to the discussion in \cite{Bekenstein:1995ju} for the case of equidistant spectrum and have been worked out in detail in \cite{Ansari:2006vg}.   Because of the uncertainty principle $\Delta E
\Delta t \sim \hbar$, one can substitute the variation of time sequence between emissions into the time uncertainty and provide the frequency width, which is found to be of the order of a thousandth of the frequency scale $\omega_o$. This shows that
the spectrum lines are indeed very narrow and the various black hole
lines of one generation are unlikely to overlap.

The intensity of all harmonic frequencies in the range of frequencies up to $10 \omega_o$ is plotted in
Fig. (\ref{fig2}). Each generation is denoted in a color. The dashed lines indicate the envelops function of radiation intensity at each generation. In this plot, one can see the first generation amplifies some harmonic frequencies whose first few frequencies have the highest intensity. The higher generations appear in harmonic lines within the lower envelopes.  In an ``observational'' spectroscopy of a black hole radiation one can see the first few harmonics of the first few generation. The inset of Fig. (\ref{fig2}) presents the intensity envelop of the first few generations in the log scale. The fist generation is the topmost curve and the second generation is the one below that. In the inset, as the plots from top to bottom denote generations from the smallest gap between level to higher gaps, respectively.

As a consistency check let us briefly study the consistency of this derivation we offered in Eq. (\ref{eq. I}) with the Bekenstein black hole entropy. The distribution of the number of quanta emitted from a black body radiation. The probability of one emission of frequency $\omega^*$ is Boltzmann-like; $\pi_{\omega^*} = B \exp(-\frac{\hbar \omega^*}{kT})$ where $B$ is normalization factor $B = \sum_{\omega} \pi_{\omega}$.

Successive emissions occur independently and therefore the probability of a $j$ dimensional
sequence with $p$ emissions of the frequency $\omega^*$ is

\beq \label{eq. P photon}
P(p, \omega_{f^*}(\zeta^*)|j)= \left(^j_p\right) (\pi_{\omega^*})^k \prod_{i}^{j-p} \sum_{\omega_i
\neq \omega*} \pi_{\omega_i}.
\eeq

The last summation term can be
replaced from the normalization relation by $B-\pi_{\omega^*}$. This probability determines the entropy of the sequence of radiation: $P \sim \exp(-S )$, where $S$ is the photon emission entropy.

A black hole is hot and the thermal character of the radiation is entirely due to the degeneracy of the levels, the same degeneracy (\ref{eq deg})
that becomes manifest as black hole entropy. Therefore its radiation is characterized by Planck's black body radiation with a temperature that matches the black hole temperature (after the appropriate adjustment of the Barbero-Immirzi parameter).  This allows one to replace $\exp(-\frac{\hbar \omega^*}{kT})$ in the definition of $\pi_{\omega^*}$ in Eq. (\ref{eq. P photon}) with its geometrical analogous probability $g_1(\zmin)^{-f^*\sqrt{\zeta^*/\zmin}}$.    One can explicitly determine this is the probability of a $j$- dimensional
sequence with $p$ emissions of the frequency $\omega^*$ in black holes as a power law function of the harmonic mode numbers $ P \sim g_1(\zmin)^A$.  This gives rise to a physical black hole entropy equivalent to Eq. (\ref{eq. Sbh}) that provides the self-consistency of our method. It is important to emphasize  this analogy does not hold perfectly if one does not consider both the population and the transition weights in the derivation.

\section{Conclusion}

A large class of semiclassical analysis and quantum gravity theories predict an unevenly spaced discrete spectrum of area for a black hole horizon area. In the case this spectrum of area contains a collection of unbounded evenly spaced subsets, similar to the case where $A \sim \sqrt{n}$, recent studies predict a large amplification of a few frequencies due to the horizon fluctuations.

In this paper we studied the quantum amplification effect by counting all emitted photons, which has been unnoticed in previous studies.  We derived the intensity scales with the harmonic quantum numbers as a power law  and unlike previous studies it is independent of the Barbero-Immirzi parameter.  

A black hole amplifies the quanta of its horizon area such that a few of them become ``observable'' in the macroscopic scales.  The exact spectrum  of these lights reveals the underlying quantum gravity signature on the quantization of area.  The highly amplified flashes of light are only a few in number about the range of frequencies $\omega/ \omega_o < 3$ in an unevenly spaced fashion.   In a typical black hole of mass $10^{-18}$ solar mass the frequency scale $\omega_o$ is in the range of 10 keV.  These modes are  even brighter in larger holes, although their frequencies become smaller.

   The different nature of this radiation from the formal Hawking radiation and the amplification effect on a few frequencies by the hole makes it possible to search for this unevenly spaced spectrum in observations.

\section{Acknowledgement}
The author thanks Jacob Bekenstein and Viatcheslav Mukhanov for their detailed comments on the first version of this manuscript. He is also thankful  to Martin Bojowald for a clarifying discussion on the properties of information flow in quantum horizons. This research was supported by NSERC Canada through Ashwin Nayak.

\end{document}